**Enforcing public data archiving policies in academic publishing: A study of ecology journals**

Dan Sholler
University of California, Berkeley
Berkeley, CA
dsholler@berkeley.edu

Karthik Ram
University of California, Berkeley
Berkeley, CA

Carl Boettiger
University of California, Berkeley
Berkeley, CA

Daniel S. Katz
University of Illinois Urbana-Champaign
Urbana, IL

**Abstract**

To improve the quality and efficiency of research, groups within the scientific community seek to exploit the value of data sharing. Funders, institutions, and specialist organizations are developing and implementing strategies to encourage or mandate data sharing within and across disciplines, with varying degrees of success. Academic journals in ecology and evolution have adopted several types of public data archiving policies requiring authors to make data underlying scholarly manuscripts freely available. Yet anecdotes from the community and studies evaluating data availability suggest that these policies have not obtained the desired effects, both in terms of quantity and quality of available datasets. We conducted a qualitative, interview-based study with journal editorial staff and other stakeholders in the academic publishing process to examine how journals enforce data archiving policies. We specifically sought to establish who editors and other stakeholders perceive as responsible for ensuring data completeness and quality in the peer review process. Our analysis revealed little consensus with regard to how data archiving policies should be enforced and who should hold authors accountable for dataset submissions. Themes in interviewee responses included hopefulness that reviewers would take the initiative to review datasets and trust in authors to ensure the completeness and quality of their datasets. We highlight problematic aspects of these thematic responses and offer potential starting points for improvement of the public data archiving process.

# 1. INTRODUCTION

The value of open data in the scientific discovery process is well-documented (Bowker, 2001; Hilgartner, 2013; Leonelli, 2013; GEO, 2015). As Michael Nielsen (2011) wrote in *Reinventing discovery: The new era of networked science* (p. 108), "Scientists in many fields are collaborating online to create enormous databases that map out the structure of the universe, the world's climate, the world's oceans, human languages, and even all the species of life." Sharing data, as Nielsen and others (e.g., *The Royal Society,* 2012) note, increases the speed and enhances the quality of scientific discovery. In some cases, creating the "enormous databases" to facilitate improved science is a direct result of answering scientific questions: No one astronomer, for example, can build and deploy the tools necessary to survey distant galaxies without direct coordination and collaboration. In other words, sometimes shared databases emerge out of necessity. Other cases require coordination of small-scale projects that could, in theory, exist as standalone pursuits without sharing data; instead, researchers recognize some value−whether scientific, legal, moral, or other−in sharing datasets.

Coordination of data sharing efforts in the latter cases relies on a number of stakeholders. Funding agencies, for example, might seek to streamline their efforts by requiring data to be shared and preventing costly re-collection. Funders have both incentives and enforcement mechanisms readily available (i.e., "carrots and sticks") (Couture et al., 2018; Diekema et al., 2014). Other stakeholders, including scientific journals, manage a delicate balance between incentives and enforcement. These journals are increasingly requiring researchers to make datasets associated with manuscripts available, often by establishing public data archiving (PDA) policies (Roche et al., 2015). PDA policies illustrate journals' recognition of data archiving as an essential step in the research process (Whitlock et al., 2010; Vines et al., 2013), yet the appropriate mechanisms for managing the data archiving process are, to date, undefined.

Journals have several motivations for instituting PDA policies and developing appropriate strategies for incentivizing and enforcing compliance. In principle, policies requiring authors to publish the datasets underpinning analyses in their manuscripts facilitate scrutinization, reproduction, and replication of studies (Bloom et al., 2014; Goecks et al., 2010). The resulting transparency can increase public trust in science (Beardsley, 2010; Duke and Porter, 2013; South and Duke, 2010) and, by extension, enhance the reputation of the journal. Journals may also view PDA as a way to increase citations, to provide other researchers interested in the same or similar phenomena with resources, and to provide valuable objects of collaboration (Borgman, 2007; Edwards et al., 2011). Furthermore, PDA policies aid in ensuring the sustainability and quality of scientific data. Without adequate PDA, the short and long-term sustainability of research data diminishes (Kaye and Hawkins, 2014; Vines et al., 2014). Incorporating review of datasets into the publishing process can help to avert some of what Leonelli (2014: 1) refers to as

"difficulties caused by the lack of adequate curation for the vast majority of data in the life sciences."

Publishers have implemented PDA policies in various journals across scientific disciplines (see Appendix A for a list of examples in the biological sciences) and many different types of policies have emerged. In general, PDA policies fall on a spectrum from those only requiring authors to make data "available upon request" (e.g., via email from an interested party) to requiring authors to deposit datasets in specific repositories housing specific types of data (e.g., *GenBank*, the universal choice for genome sequencing data). Each approach has benefits and drawbacks that require editorial staff to balance incentives and enforcement. As we discuss in the next section, many journals have moved beyond "available upon request" policies and instead fall somewhere between voluntary dataset contribution and mandated PDA. For example, some journals require authors to write data availability statements (brief attestations to where the data are located) and allow authors to choose from a variety of repositories to house their data.

Moving beyond "available upon request" policies served as a consensus step toward realizing the potential of open data in science. As Michener describes, factors such as the availability of technical infrastructure for data sharing and funder policies requiring sharing drove these changes in norms (Michener, 2015). However, the effectiveness of PDA policies for enabling reproduction, replication, and data reuse remains questionable. For example, Roche et al. (2015: 1) found that 56% of published datasets related to manuscripts in top ecology journals were incomplete, and 64% "were archived in a way that partially or entirely prevented reuse."

This study investigates why PDA policies that go beyond "available upon request" may not be proving effective in realizing the goals of open data efforts. One purported reason for failing to build collections of reusable datasets is that journals lack appropriate enforcement mechanisms and incentive structures to ensure that published datasets are complete and high quality (Costello et al., 2013; Mayernik, 2017). We sought to build on this idea and explain the mechanisms by which journals enforce PDA policies by examining the roles of stakeholders in the PDA process and identifying problematic aspects of the process. We do so by reviewing the state of PDA policies in a broadly-defined biology discipline—ecology and evolution—and identifying some of the unique challenges of sharing data in this area of research. We chose ecology and evolution because of its scope (i.e., the range of biological and other sciences it covers) and its long history. We begin by describing the types of PDA policies currently in place. We then describe our interview-based approach to examining how PDA policies are enforced in ecology journals, present the themes we identified in interview responses, and contextualize the findings in the ongoing discussions about data and related research artifact sharing in science.

# 2.     JOURNAL DATA PUBLICATION POLICIES IN ECOLOGY: CURRENT STATUS

In Star and Ruhleder's (1996) foundational piece on cyberinfrastructures, they emphasized that infrastructures such as those supporting open data do not grow *de novo*. Infrastructures such as those for sharing data are shaped by existing tools, methods, and practices; therefore, studying infrastructures of any kind requires contextualizing the study in a situated way (Jackson et al., 2007; Schrock and Shaffer, 2017; Tilson et al., 2010). In other words, studying open data infrastructure development and use should recognize and account for the diversity of stakeholders and knowledge-making processes involved in the scientific discovery process (Bowker, 2006; Edwards et al., 2011; Hine, 2006; Jasanoff, 2004; Kitchin, 2014; Knorr Cetina, 1998). Adopting this approach, we examined the enforcement of PDA policies by journals within one discipline.

## 2.1 Data sharing in ecology and evolution: A brief history

Data sharing in ecology predates the recent shift toward open data. As Michener and Jones (2012) described, large projects in ecology and evolution required data sharing as early as the 1980s with the advent of programs like the Long Term Ecological Research (LTER) Network. LTER has grown in size and shifted in structure over its 40-year history, but has always managed data from at least 15 ongoing research programs spanning forest, grassland-agriculture, tundra, coastal, freshwater, marine, and urban ecosystems (LTER, 2018). Examples of data sharing in ecology and evolution can be found even earlier than the 1980s. For example, the International Biological Program ran from 1964 to 1974 and served as a foundational approach for collaborative, large-scale ecosystem science (see Coleman (2010, pp. 15-89) for a detailed account of its emergence and impact).

Recently, subdisciplines and interest areas in ecological research have developed their own infrastructures for sharing data. For example, the Global Biodiversity Information Facility houses hundreds of datasets and continues to grow in size and scope. However, as Costello et al. (2013) pointed out, over two-thirds of the datasets have been provided by government organizations rather than from academics. This finding is surprising given that, according to Ware and Mabe (2015) and Costello et al. (2013), academic researchers publish 75% of all scientific papers across scientific disciplines. Researchers from various subdisciplines have called for increased contributions from the academic community and have offered mechanisms by which the community might review and improve datasets prior to publishing (Chavan and Ingwersen, 2009; Costello, 2009; Costello et al., 2013; Michener, 2015; Piwowar et al., 2007).

A turning point in the effort to institute PDA policies came in 2008 with the launch of *Dryad*, a data repository born out of a workshop held by the National Evolutionary Synthesis Center (Dryad, 2007). The workshop, it seems, played a similar role in ecology as the historic 1996 genomics conference played in ushering in a new era of genomics data sharing via the Bermuda Agreement (see Nielsen, 2011: 149-162). Conveners of the workshop conceptualized a repository for small scientific communities to aid in the preservation and sharing of datasets from evolution studies. The workshop and ensuing discussions prompted researchers to raise issues with data sharing: What is the appropriate level of data to archive—data underlying figures, raw datasets, or something in between? How can we expect researchers to contribute data before they have completed all of the analyses they wish to conduct?

As *Dryad* grew in size and organizational structure, in part as a result of funding from the National Science Foundation, it also grew in scope. Journal editors in the ecology and evolution fields began to take note of the potential for integrating *Dryad* into the manuscript publication process, as evidenced by editorial pieces in *Nature News* (Nelson, 2009), *The Journal of Evolutionary Biology* (Moore et al., 2010), *The American Naturalist* (Whitlock et al., 2010), and *Trends in Ecology and Evolution* (Whitlock, 2011). These editorials often cited the success of *GenBank* in prompting genomics researchers to share genome sequencing data. Moore et al. (2010: 659) articulated the need for an ecological version of *GenBank* in an op-ed in *The Journal of Evolutionary Biology*:

> *"The example of GenBank shows the value of the availability of data for all of these reasons. The modern synthetic use of DNA sequence data would not be possible without the near-universal use of GenBank as a public archive. Moreover, GenBank would not be nearly as complete as it is without the communal decision to archive all DNA sequence data, a decision initially introduced by journals."*

In contrast to LTER—which required only those researchers funded by the LTER program to make data publicly available—journals could mandate that all authors deposit data. *Dryad* enabled editors and other editorial staff to provide a viable option to authors rather than suggesting authors make data available upon reasonable request. The emergence of technical infrastructure via repositories such as *Dryad* enabled journals to develop a range of PDA policies, some of which are discussed in the next section.

## 2.2 Current status: PDA policies and technical infrastructure

Journals, perhaps partially in response to calls for data contributions from their manuscripts and the availability of data repositories, are rapidly adopting and implementing PDA policies. Figure 1 places these policies on a spectrum, from no policy at all (rare), to requiring a data availability statement (most common), up to a full data review process resulting in peer-reviewed datasets (rare).

**Figure 1. Types of PDA Policies**

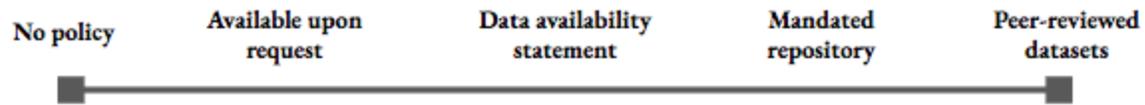

### 2.2.1 *Available upon request*

The first journals to implement PDA policies required all authors to make data available upon reasonable request. In other words, editors mandated that authors share their data should someone contact them to examine or reuse the datasets underlying manuscripts. These policies tended to be non-specific with regard to the form and amount of data to be shared (e.g., data underlying figures in the manuscript vs. raw data). "Available upon request" policies served as an initial, consensus-driven step toward making datasets widely available.

Researchers have since tested the efficacy of "available upon request" policies and levied criticisms against the policies' ability to reach desired outcomes, backed by evidence that authors often do not respond to such requests or deny the requester access. Stodden et al. (2018), for example, conducted a study in which the authors requested data and code from a random sample of 204 manuscripts in *Science* and attempted to replicate the analyses. The authors received responses with data and/or code from 44% of the sample and had sufficient information to reproduce the findings for 26% of the manuscripts. The paper concludes that available upon request policies are "an improvement over no policy," but do not fully support availability or reproducibility.

### 2.2.2 *Data availability statement*

Many journals in ecology and evolution have moved beyond available upon request policies and have made strides toward ensuring the long-term availability of research data. Like journals in many other disciplines, ecology journals have favored data availability statements as the preferred requirement for PDA. A data availability statement requires authors to write a section detailing where the data can be found. In general, data availability statements can point to a number of places, including supporting information files and public repositories. Editorial policies at journals such as *PLOS* (2014), for example, strongly suggest that public, discipline-specific repositories be used. However, exceptions are available and allow authors to, at minimum:

> ... specify "Data available on request" and identify the group
> to which requests should be submitted (e.g., a named data
> access committee or named ethics committee). The reasons
> for restrictions on public data deposition must also be
> specified. Note that it is not acceptable for the authors to be
> the sole named individuals responsible for ensuring data
> access.

### 2.2.3 Mandated Repository

The availability of repositories like *Dryad* enabled journals to move toward mandating that researchers archive data in a public or semi-public outlet. The Joint Data Archiving Policy (JDAP) is perhaps the exemplar of this type of policy. JDAP is a general policy that journals may use as part of their editorial policies. The template text, as presented on *Dryad's* website, reads:

> *[Journal] requires, as a condition for publication, that data supporting the results in the paper should be archived in an appropriate public archive, such as [list of approved archives here]. Data are important products of the scientific enterprise, and they should be preserved and usable for decades in the future. Authors may elect to have the data publicly available at time of publication, or, if the technology of the archive allows, may opt to embargo access to the data for a period up to a year after publication. Exceptions may be granted at the discretion of the editor, especially for sensitive information such as human subject data or the location of endangered species.*

Michener (2015) identified 16 repositories holding ecological data, some of which provide authors with the ability to submit data. Services such as Re3data.org catalog repository options for authors in ecology and evolution and other disciplines. Additional services such as Science Europe's Framework for Discipline-specific Research Data Management aid researchers in selecting potential homes for datasets to adhere to PDA policies set forth by journals and other governing bodies (Science Europe, 2018).

Ambitious examples of PDA policies extending beyond JDAP also exist in ecology and evolution. These policies specify the level of data authors should submit and, in some cases, which repositories should be used. Specific levels of data vary. *PLOS Biology*, for example, requests that authors submit the data underlying the findings in the paper. Other journals specify which components of the paper should be supported with data submissions. *Ecological Applications* (2018), for example, maintains the following PDA policy:

> *Archived data should be sufficiently complete so that subsequent users can repeat tables, graphs, and statistical analyses reported in the original publication, and derive summary statistics for new or meta- analyses. Thus, the normal resolution of the data that are archived will be at the level of individual observations.*

Journals requiring specific repositories often do so for particular types of data, rather than as a general rule. For example, the journal *Evolution* requires that "DNA sequence data must be submitted to *GenBank* and phylogenetic data to *TreeBASE*" (Fairbairn, 2011: 1).

### 2.2.4 *Peer-reviewed datasets*

The least common policy among those presented here are policies resulting in peer reviewed datasets. The journals with such policies tend to be specialized instantiations of a larger journal. In other words, these journals often specialize in data publications and therefore may not have the visibility or readership traditional journals enjoy in the ecology and evolution research communities. *Nature's Scientific Data* journal, for example, offers authors the opportunity to submit, receive reviews on, and publish "Data Descriptors." Data Descriptors are manuscripts explaining the collection and format of unique datasets (Scientific Data, n.d.). *Ecology's* Ecological Archives provides a similar example (ESA, n.d.). These types of publications achieve data review rigor unseen in traditional journals.

### 2.2.5 *Research Questions*

Journals that have such data review processes produce high-quality datasets enabling future use by the scientific community. However, the degree to which traditional manuscripts receive this type of attention from reviewers was, prior to this study, unclear. We sought to illuminate the process by which traditional journals enforce PDA policies and ensure the quality of published datasets by interviewing editors at widely-read ecology and evolution outlets. The following research questions guided our study of the PDA enforcement and data review processes:

> RQ1: *How do journals develop, implement, incentivize, and enforce PDA policies?*

> RQ2: *Who do editors perceive as responsible for enforcing PDA policies and reviewing datasets? What roles do they assign to each participant in the peer review process?*

## 3.   METHODS

We conducted a qualitative, interview-based study of how journals enforce data publication policies. We selected ten journals that publish research in ecology and evolution, all of which have adopted either JDAP or their own PDA policies. The study data were collected by conducting 20 semi-structured interviews with journal editors (editors-in-chief, associate/assistant/subject editors, and data editors), data repository staff, and journal/publisher staff. We followed qualitative data analysis techniques first developed by Miles and Huberman to identify themes in the responses to our semi-structured interview questions (Miles and Huberman, 1994), focusing explicitly on editors' and staff's perceived roles of each stakeholder in enforcing PDA policies.

### 3.1 Data Collection

We employed three semi-structured interview protocols to provide the qualitative data necessary to assess how journals develop and enforce data sharing policies. We used one protocol to guide our interviews with editors and editors-in-chief; one protocol for associate and assistant-level editors; and one protocol for repository/journal staff. We interviewed different types of stakeholders in the PDA policy enforcement process following the advice of Weiss, who emphasized the importance of capturing diverse perspectives when constructing an interpretation of an organizational process (Weiss, 1995).

In developing the protocols, we followed guidelines set by Spradley (1979) for conducting semi-structured interviews. Semi-structured interviewing involves the use of a guiding protocol (i.e., a general set of questions that the researcher can adapt to the ongoing conversation) for the interviews rather than a strict protocol or no protocol at all. The semi-structured interview allows the researcher to explore emergent ideas that may provide new insights into the phenomenon of interest, whereas a strict protocol might preclude the interviewee from expressing what is important to his or her thinking on a given topic. Interview protocols contain common questions so that responses can be compared across informants (Bernard, 1988; Yin, 1994), a feature that is vital to identifying and validating themes in the data (Corbin and Strauss, 1990; Emerson et al., 1995). In this act of comparing and contrasting, the researcher can ensure that the sentiments and experiences of the informants are not idiosyncratic, but are in some way thematic and/or related to one another.

Similarly, we structured questions in a way that elicited responses related to specific incidents or tasks that are central to the phenomenon of interest. Spradley (1979) also noted that questions are more effective when they are designed with a working knowledge of the interviewees' work than when designed from an external perspective. Accordingly, we leveraged our knowledge of the publishing process in ecology and evolution, of commonly-held beliefs about the barriers to data sharing, of technologies that enable data sharing, and of ongoing discourses about open data initiatives to inform our guiding questions.

The length of interviews varied and ranged from 38 minutes to 1 hour and 26 minutes. Interviews were held by webconferencing technologies, via phone, and in person; audio recorded; and transcribed by the corresponding author. The audio files and transcripts were stored on a secure institutional server according to human subjects policies at the host institution. Transcription included making notes about details such as the inflection of the informant's voice or interruptions experienced during the interview. Table 1 summarizes the number of interviews by stakeholder group.

**Table 1. Interviews with editors, AEs, and repository/journal staff.**

| Role | Interviews |
|---|---|
| Editors/Editors-in-Chief | 8 |
| Associate or Assistant-level Editors | 6 |
| Repository/Publisher Staff | 6 |
| **Total** | **20** |

## 3.2 Data Analysis

We employed qualitative coding techniques to analyze interview data. Interview transcripts were analyzed using Atlas.ti qualitative coding software. Following the work of Corbin and Strauss (1990), Emerson et al., (1995), and Miles and Huberman (1994), the analysis was interpretive and iterative. The analysis began with repeated readings of interview transcripts and continued with readings until we were familiar with all of the actors, terminology, policies, and interaction scenarios contained in the data. The next step, open coding, involved annotating the interview transcripts with initial interpretations of the data based on the initial repeated readings of transcripts. A simple example is: Repeated readings called our attention to a universal penalty for noncompliance with PDA policies—refusal to publish the paper—and we marked the transcript by applying the code "penalty - refusal to publish" wherever interviewees discussed the penalty and how it was applied. Like this example, the codes developed in open coding tend to be general and descriptive. Following Corbin and Strauss (1990), codes were then refined, conflated, or made more granular throughout the open coding process. We then used several other forms of qualitative coding to deepen the analysis. For instance, selective coding involved looking closely for additional instances of a particular concept or theme. Axial coding included relating codes to one another by breaking single codes down into multiple, related codes or recognizing co-occurrence of two or more themes and conflating them. Glaser's constant comparison was employed throughout (Glaser, 1965). Constant comparison refers to the process of comparing new instances of a code to its previous uses to ensure that there is thematic consistency. We present the themes generated from the analysis below.

## 4.    FINDINGS

The journals we studied implemented formal data sharing policies between 2013 and 2015. At the time of implementation, all ten journals had existing guidelines in place encouraging authors to share data associated with their manuscripts. The increasing availability of technical infrastructure to support data sharing in ecology (e.g., growth in the number of repositories) and community criticism of "available upon request" policies prompted the expansion of data sharing policies. Editors and repository staff reported that transitioning from voluntary submission to mandated data archiving did not appear to be an unreasonable extension of existing practice.

   All journals opted for a data availability statement approach, with varying levels of and approaches to enforcement. Sometimes, reviewers were asked to check that a link to data was present; others were encouraged to click on the link and ensure that datasets were in a reasonable repository. According to editors, no journals asked reviewers, AEs, or editors to systematically scrutinize the published dataset. Editors viewed AEs and reviewers as jointly responsible for enforcing compliance with the data availability statement and verifying that the data exist at the location provided by the author. AEs relied on reviewers to flag potential quality issues early in the review process. However, editors and AEs alike repeatedly emphasized that the authors were primarily responsible for ensuring the quality of their datasets. Below, we describe the implementation of PDA policies, the mechanisms of their enforcement and quality assurance, and editors' plans going forward.

### 4.1 Developing Data Publication Policies: "available upon request" to "available"

   All of the journals we studied had implemented or were beginning to implement PDA policies at the time of our interviews with editors. A common theme in participant accounts of the implementation of data sharing policies was that mandated sharing emerged through bottom up changes to practice within the ecology and evolution community. Editors commonly made statements such as, "It's not actually hugely different from the previous policy" because most journals "already operated with a kind of a policy expecting that data would be shared upon reasonable request." Similarly, one editor stated, "we had a data policy that very very strongly encouraged the archiving—well, it was the same policy, but it was optional instead of mandatory." Indeed, in almost all of the journals, transitioning to data availability statements was a matter of moving from voluntary contributions to mandatory archiving, with little additional action required. Editors reported little pushback from authors regarding the extension of data sharing policies. As one editor of a journal requiring a data availability statement recounted,

> We've gotten shockingly little pushback on our open data policy. We literally had one author write and say, "My career depends on my exclusive access to these data." In 2 years … I want to say that we've really only ever in two years have one author say that they did not want to make their data publicly available because they hadn't finished writing all the papers they wanted to. Literally, there's n of 1 on that. We've had a number of authors who've had to work with us to come up with a compromise around data that could not be publicly shared. Either because it involved endangered species or human subjects, or because it involved data that was obtained through, like, a commercial fisheries program.

One editor reported slight initial resistance, but that resistance was based on a miscommunication of the policy. The journal published a blog post outlining the new policy and received critical responses on social media and via email. Authors were concerned that the journal was asking authors to submit all raw data from their research projects. The journal issued several additional blog posts clarifying that the journal was only requesting data underlying figures in manuscripts. According to the editor and one AE at the journal, the clarification alleviated author concerns.

Asked to speculate as to why they encountered little pushback, editors commonly pointed to three factors: ecologists' history of sharing data in projects that predate journal data sharing policies; a new generation of researchers "growing up" with the expectation that data would be shared; and the ethos of the ecology and evolution field. Editors-in-chief eagerly discussed ecology's history of data sharing and pointed to it as evidence that the discipline was well-positioned to adhere to data sharing policies. Editors-in-chief tended either to cite their own involvement in data sharing projects before the "open data movement" began or to describe examples of large, collaborative ecological projects. As one editor reported, he had firsthand experience with collaborative projects requiring data sharing as early as the 1980s:

> I was funded on a [federal agency] project, I think we were selected in 1984. And that was the first project that [federal agency] enforced its earth science data policy for. And so I've worked in the context of open data since I was a postdoc. And I'm very much in the context of the [federal agency] open data policy, which applies not only to [redacted agency name], but to all of the field projects and

> *research that the agency supports. And so, because of that, I've been involved not only in collaborative open data projects – not as a practitioner by the way – but also I've seen the evolution of the data systems that support those open data policies. So my first project, the [redacted] project in the late 1980s, actually had a data system to support and enable the team to share their information. Of course, back then, we were sending three-and-a-quarter inch floppy disks back and forth with our data on them because there was no Internet. But I've kind of seen the evolution from an earth science perspective of open data for a long time.*

Other editors had similar stories derived from working on projects such as the LTER program:

> *At least in U.S. ecology, large collaborative projects like LTER and FluxNet go back quite a ways … So community ecologists have grown up in an environment where either they themselves or they were around people who were sharing data. So I think that our data sharing activities go back further than the open data movement. And, you know, I think that the motivation that I expressed, this notion of transparency for decision support, that's something that ecologists have been dealing with for a long time in supporting decisions like the Northwest Forest Management plan, the spotted owl issue, or salmon fisheries, or coastal zone management. So I think ecologists have been working in an environment where they might have needed to make their data available for scrutiny for quite a long while.*

Interestingly, the same editors-in-chief who cited ecology's long history of data sharing also noted that "the younger crowd" had been trained and conditioned to expect data sharing. They discussed early career researchers' attitudes toward data sharing as if a shift had occurred, yet contextualized their experiences with PDA in a historical perspective. This seemingly contradictory view influenced how they expected PDA policies to be enforced in at least 5 of the journals: Editors-in-chief expected that authors would submit quality data by default ("I *think that a lot of the authors in* [our journal] *are younger, and a lot of them have grown up in the open data world*") and, perhaps more importantly, also expected that AEs (who tended to be early-to-mid career) would hold authors accountable:

> We really rely on the AE to make sure, and the reviewers, but mainly the subject matter editor, to ensure that what's archived is usable. (editor-in-chief)

> All of the quite junior people have become quite expert at fielding these types of inquiries [from authors about data sharing]. (editor-in-chief)

Some editors also explained that the lack of resistance they encountered when implementing PDA policies might relate to the ethos or underlying motivations of the field. This theme emerged in two interviews with editors-in-chief and one interview with an AE. This rationale comprised two points. The first is that ecology and evolution researchers hold views about using their research to benefit society. Citing examples such the conservation sub-discipline, editors explained:

> I'm going to speculate that ecology is a pretty societal-benefit oriented field … and we have the motivation to share data that comes from people's motivation related to conservation and other sort of benefit areas. (editor-in-chief)

Furthermore, editors speculated that ecology was in a strong position for data sharing because of the lack of profit motive:

> There's not as much private sector influence. Engineering, these types of fields, they can make big bucks off of their research products. (AE)

> We don't have the overlay of the profit motive that biomedicine has … So I think we have the same sort of fairly practical view that biomedical science does, but without the overlay of big money. And so nobody thinks they're going to get rich off their ecological data or code or anything like that. (editor-in-chief)

## 4.2 Enforcing Data Publication Policies

The lack of resistance to the policy described by the editors did not mean that authors were entirely compliant with data sharing policies, nor were editors unaware of the quality issues associated with archived datasets. In fact, editors reported that their journals did, on occasion, need to take action to ensure that authors submitted datasets upon acceptance of their manuscripts. Interestingly, the mechanism of that action varied across journals, and perceptions of whose responsibility it might be to take appropriate action garnered no concrete answers. Editors commonly cited heterogeneity of ecological data as an obstacle to establishing standard mechanisms for enforcing data sharing policies and, when probed, reported that they routinely handled compliance issues on a "case-by-case" basis.

Editors, AEs, and repository/journal staff discussed the roles of at least four key groups of stakeholders who participate in the journal PDA process and commented on how each had a hand in the enforcement of PDA policies and ensuring the quality of research data: Authors (researchers), reviewers, editors, and repository staff. Commonalities emerged in the responses interviewees gave about the role of each stakeholder in the enforcement of data policies (i.e., Reviewers, Associate/Assistant-level editors, Repositories, Authors). We focus below on how editors perceived reviewer and author roles and, in the process, discuss how editors view their own roles in enforcing PDA policies and ensuring data quality.

### 4.2.1 Role of Reviewers

Editors and associate editors reported that they do not select reviewers based on expertise in data or code; instead, editors aim for reviewers with domain expertise. Expertise sometimes includes adeptness with making data reproducible/reusable, but often does not include these skills to the same degree that we might expect to find in fields like mathematics or computer science. Journals that publish methods paper often accept manuscripts heavily focused on data and/or software code (e.g., *Methods in Ecology & Evolution*); the review process for these types of papers constitute an exception to the themes discussed below because they explicitly acknowledge datasets and/or code as a novel method and treat review accordingly.

No editors reported choosing reviewers based on their ability to review datasets, nor did they expect that a reviewer with domain expertise would be capable or willing to review datasets within their domain. Instead, editors-in-chief and AEs repeatedly expressed "hope" that reviewers would have the skills and appetite to review datasets:

> *We hope that reviewers will have flagged issues along the way if they can't access data, etcetera, as well.* (editor-in-chief)

> *So there are specific questions in the reviews* [guide for reviewers] *at* [some journals] *about whether all of the data should be available are available, so we hope that that question gets looked at externally by reviewers or the academic editor.* (AE)

Editorial staff noted that little had been done to integrate data review into the review process and guidelines for reviewers. In some cases, editors and AEs asked reviewers to read data availability statements and acknowledge that they had confirmed dataset availability. But editorial staff reported that there was little effort at their journals to build processes into peer review in order to remedy any data quality issues. The prevailing arguments were that data appropriateness for a given study was more important to the integrity of the research than issues of formatting, metadata, and other quality indicators, as this exchange demonstrates:

> INTERVIEWER: So *what do you do to build new things into the review process to combat that sort of bias or, maybe the tendency for some things to be included* [in a dataset], *others not?*
>
> ASSOCIATE/ASSISTANT-LEVEL EDITOR: *Well, I don't think there's anything explicit we do in the review process. It's just, when a paper comes in that analyzes something like this, is it, is the scope of the inference and the scope of what they're doing really relevant? For example, if you're analyzing something about body size, are there any experimental studies that have ever shown that body size is important for that? If there are, then sure, go ahead, but if there's not, well that's, you know...*

At some journals, reviewers were not required to acknowledge that they read data availability statements and ensured that data were present or high quality. Several journals we studied, and many journals in general, have free form reviews that allow reviewers to include any feedback they see fit. Editors and AEs reported that, in some cases, reviewers would use portions of free form reviews to discuss data quality issues; however, AEs in particular pointed out the rarity of such a review. "Out of every ten [reviews]," one AE explained, "maybe one will raise issues with the dataset itself. They mostly focus on what is presented in the manuscript."

In cases where the journal required authors to submit a data availability statement, but reviewers were not asked to certify the veracity of the statement, PDA policy enforcement was treated as a copy-editing process.

> *It's more of an administrative process ... We do have some minimum checks, though, things like making sure there are no instances of "data not showing"* [referring to text in the paper]. *So then we'll kind of macro-level check on what's written in the paper.*

Additionally, employees at the repositories housing the data from the top ecology journals reported a limited data review process. When asked, repository employees reported that journals provide little to no guidance in what repository staff should look for when working with authors on data publication. Repositories' internal policies dictate, to some extent, that datasets adhere to FAIR principles (Wilkinson et al., 2016). However, no editor or repository employee reported reaching an agreement to reject a manuscript or otherwise significantly delaying publication due to dataset attributes.

### 4.2.2 Role of Authors: "Trusting the authors"

Perhaps the most interesting of the themes that emerged from our analysis concerns editors' perceptions of the role of authors in PDA policy compliance and data quality assurance. In particular, editors-in-chief and AEs repeatedly mentioned "trust" for authors as a key aspect of their PDA practices. Their responses echo arguments that have been made elsewhere regarding the need for culture change, rather than policy, to drive changes in data sharing norms. Surprisingly little consideration was given to imposing further PDA policies upon authors; furthermore, several editorial staff members discussed hesitation at the idea of stricter policies for fear of driving authors to submit to other publications.

One of the factors underlying editorial staff's dependence on trusting authors was community disagreement over the appropriate "level" of data archiving. The participants we interviewed discussed their inability to reach consensus about what should be shared and how it should be shared (presumably having moved on from discussions of whether or not *something* should be archived). Regarding what should be shared, editors reflected on what they considered when developing the PDA policies and their partnerships with data repositories. When asked about whether or not repositories perform a final data check, for example, editors often reported that they could not ask repositories to do so because even reviewers and editors could not decide what level of data is archived beyond what underlies the figures in the paper:

> INTERVIEWER: *Is there anyone at* [data repository] *that's checking the data for you, to see if it's complete, if it's appropriate for the paper, and that sort of thing?*
>
> ASSOCIATE/ASSISTANT-LEVEL EDITOR: *It's completely on the author. We went through a very long period about what is the appropriate way to do that, what is the appropriate sort of level of specificity for the data, what's the appropriate level of annotation, and you know, once you start getting into those issues, and you're running on volunteers, it just blows up in your face.* [laughter] *And at*

> *some level, too, it's never really clear exactly what is the appropriate level.*

The above statement highlights some of the issues inherent in relying on volunteers to review manuscripts and associated research artifacts. Resolving the issues, then, might require engagement with and education of the broader scientific community, as some of the editors we interviewed mentioned. Engagement and education, they suggested, might begin with discussions of the appropriate expectations for authors' data submissions. Editors commonly expressed a desire to work towards agreements, even with the acknowledgment that PDA policies might not rapidly change alongside these consensus agreements:

> *I think what we'd like to be doing more is working with the community to define what the data sharing expectations are in individual sub fields and disciplines and for different specific data types. So you might look at something like fMRI and try and understand which pieces of the data output are required and which format they're best put in, and basically every subfield, you might look at electrophysiology data, or it might be imaging data, which is another very complex problem because it can be such big amounts of data. So I think what we'd like to do is work with the community to better define and understand what the expectations are and should be that is accepted by them. Because what doesn't work well is if a publisher presents an edict to a community which they haven't been involved in discussing. [editor-in-chief]*

Similarly, an associate/assistant-level editor gave the example of long-term data collection efforts when asked about how the journal, authors, and repository influence decisions about embargo periods on datasets. The editorial staff member noted that researchers did not want to "short-circuit" their future studies by publishing data and that the journal trusted the authors to make the appropriate decision:

> ASSOCIATE/ASSISTANT-LEVEL EDITOR: *And their* [staff at the data repository] *opinion is, it's up to the editor and journal and the authors to come to an agreement about what that embargo period should be.*
>
> INTERVIEWER: *But do authors have a direct line to* [the repository] *to ask them for an exception? Do you allow that? Or you come to an agreement with them?*

> ASSOCIATE/ASSISTANT-LEVEL EDITOR: *We come to an agreement with the authors, and then the authors pick. But we assume everybody's upstanding, and all we check is that data's been deposited.*

Trust for authors played a pronounced role even when evaluating the dataset (or, in this example, code) was not an option in the first place. In the following scenario proffered by an editor-in-chief, for example:

> *Sometimes the data might be something like a MATLAB script and we can't actually open it. You know, sometimes you're dealing with a proprietary file type that we can't verify. But, we sort of take that trust element of, we've got that file, we probably think it's reasonable. If there's a problem with it flagged down the line, we'd need to pursue it, but for now, we'll take that on trust.* (editor-in-chief)

### 4.2.4 *Role of Penalty.*

Editors universally reported in all interviews that the penalty for noncompliance with data sharing policies was refusal to publish the paper. Specifically, enforcement would be levied between the time reviewers accepted the paper and the time the article was assigned a DOI:

> *Well the enforcement is really pretty simple—they deposit or they don't publish. And so it's part of our acceptance checklist that we won't put the paper into production unless a staff member has gone, checked, and verified that the data are in fact present where the paper claims they are. To the extent that there's enforcement, it falls to the AE backed up by me.* (editor-in-chief)

The above scenario left little opportunity or incentive for reviewers to review the dataset prior to a decision on the manuscript. When asked about this issue, editors clarified that they expected issues to be identified and remedied before the manuscript decision step. Final checks before publication served as a formality of sorts:

> INTERVIEWER: *Ok, so that would come at the time that manuscript had actually been accepted—*

EDITOR-IN-CHIEF: *We would try to flag that earlier [chuckles], and we'd hope that that would surface earlier, but failing earlier, there's a final step that happens to make sure that the data have been made available. Ideally someone's looked at it, you know, and assessed it, whether the in-house person or the external person.*

*Well,* [data repository name] *actually sets up a link, so when we go to copy edit a paper, if there should be data in* [repository name], [repository name] *has already sent a link to be published with the paper, to say, "Here's the URL if you want the data." So we know exactly if it has been deposited or not, based on whether we got that URL or not.* (associate/assistant-level editor)

When probed, editors responded that they rarely—if ever—were forced to reject a paper on the grounds of insufficient or missing datasets. The interviewer asked for particular examples of the "few" instances, but editors could not recall those instances. Three associate editors were able to describe situations in which they had to delay the publication of a paper until an appropriate link was provided. Two of these cases involved authors who had simply not submitted a link to the dataset; AEs explained that when they received the link, they clicked to ensure that the link worked and that it directed to an appropriate repository. In the other case, the AE negotiated release of a subset of the data to protect the location of endangered species.

The dearth of cases in which papers were rejected for data archiving issues does not provide sufficient evidence that rejection fails to provide an appropriate enforcement mechanism. However, when considered alongside findings above regarding reliance on reviewers, trust for authors, and previous studies about the low quality of archived data (e.g., Roche et al., 2015), the finding does indicate that the timing of the enforcement may be problematic. In other words, authors may feel compelled to submit a dataset in some form once the manuscript has been conditionally accepted, but may not be as attentive to the completeness or quality of the dataset as they might be if it were reviewed during the peer review process. When the issue of timing was raised in interviews, five editors cited resource constriction as a limitation (e.g., "Our reviewers are already pressed for time" and "It would be too costly to decouple the review process and have dedicated staff [to review datasets]."

## 4.3 Plans going forward.

In addition to the themes about enforcement presented above, we also found commonalities in how editors discussed their journals' plans going forward. As noted earlier, many of the editors were hesitant to make PDA policies increasingly strict—i.e., by instituting peer review processes for ensuring dataset quality. Instead, almost all of the editors we interviewed expressed optimism that cultural change would drive improvements in quality of data:

> *Where, in the beginning, we had some resistance from our academic editors and reviewers about looking at the data, now we see more people who are requesting things upfront, saying "Where are the data? I'm not going to look at this further until I get the data." And so I like to think that we've seen change happen and that more people are caught unaware by the policy, which is really, it's very anecdotal, but it's the most encouraging thing I can really say in terms of what's changed since we started doing it. I don't know how many people actually go and read the data policy, but that's another issue ... they want to be able to look at the data that underpin the study before they'll review, or in order to review the paper.* (editor-in-chief)

The interviewer asked the editors who expressed optimism about cultural or community-driven change to elaborate on their reasoning. An emergent theme related to the increasing computational competence of researchers in the ecology and evolution fields. Particularly, editors pointed to the fields' engagement with software development, which they viewed as closely linked to engagement with PDA. Whether this perceived correlation holds weight remains to be seen; however, the conflation of data and software issues recurred throughout our interviews. Interestingly, editors' optimism regarding community engagement with software development in the research process did not translate to optimism about software review and sharing. Instead, editors were hesitant about extending PDA policies to explicitly extend to software code, perhaps as a result of their experiences with developing and implementing PDA policies and the assumed link between the two practices. As two editors described,

> *I think we're probably slightly shy of going full mandate direction after the data policy because it has proven so difficult to resolve the actual enforcement parts. And I think, clearly [our journal] is in a unique position, being more mission-driven and a not-for-profit, which allows us to do things that other publishers wouldn't do in the pursuit of*

> *greater openness. But I'm sure that whatever we come up with will be the subject of a lot of long and hard and tough discussions and back and forth about, it's a fine line between asking for more and ensuring that authors can actually comply, can and will comply with what you're asking of them. So it will be a complex discussion, I'm sure.* (editor-in-chief)

> *So we've put off the requirement for archiving code and scripts partly just because we wanted to see how the data policy worked.* (editor-in-chief)

Some journals, though, had decided to move on to software sharing policies and were embracing a mandate or policies similar to existing data sharing policies. Still, editors at these journals expressed concern about the difficulty they might encounter when instituting such policies.

> *I think that's going to be harder. You know, having been in the "computer-aided ecology biz" for a long time, I'm expecting to get screenshots of MATLAB procedures and horrible Python code that even the author can't read anymore, and I don't know what we're going to do about that. Because in some sense, you can't push too hard because if they go back and rewrite the code or clean it up, then they might actually change it. And so, we've talked about this across the journals.* [editor-in-chief]

Editors referenced discussions across journals, and even across subdisciplines, in three interviews. However, they did not indicate that they had discussed data and software review procedures with editors of journals in other disciplines. We discuss the potential for disciplines to learn from one another, among other potential avenues for improving PDA policies and processes, in the next section.

## 5. DISCUSSION

The findings presented above are encouraging in that editorial staff at widely-read ecology and evolution journals are reflecting upon the successes and shortcomings of PDA policies. Furthermore, the staff's recognition that technical infrastructure for data sharing exists and no longer limits authors in publishing datasets promises to shape the negotiations between journals and authors in favor of increased sharing. However, our analysis revealed several remaining issues that illustrate how going beyond "available upon request" PDA policies may not be sufficient in producing desired outcomes.

The findings highlight how incentivizing and enforcing PDA policies has proved difficult for journals in ecology and evolution because editors and other journal staff rely heavily upon hope in reviewers' attention to datasets and trust in authors to submit appropriate data. This relational approach risks reproducing the issues that limited data sharing prior to PDA policy implementation. Furthermore, reliance on reviewers and authors comes at the expense of the development of new or extended processes for ensuring data submission and dataset quality. Below, we contextualize the findings in previous studies and offer suggestions for ways to remedy remaining issues. Our primary argument is that journals might consider changes to their existing processes and adapting the processes of other organizations rather than inventing new processes altogether. Specifically, journals might be particularly amenable to processes that resemble and/or extend existing peer review processes.

The primary, remaining concern for the future of PDA policies is that data submissions, without scrutiny, will fail to reach "community standards" as defined by guidelines such as the FAIR principles (Wilkinson et al., 2016). Indeed, Roche et al. (2015) discussed the paucity of high-quality datasets emanating from ecology and evolution journals as compared to datasets published in accordance with funder policies. Given the results of our study, the lack of quality in datasets comes as no surprise: The journals we studied, although optimistic about reviewers' and authors' behaviors, did very little to equip reviewers and authors with the tools and processes to evaluate and refine archived datasets. Below, we discuss some possibilities for data review processes to be implemented in the peer review process.

## 5.1 Applying strategies from data curation initiatives

Ecologists have called for peer review of datasets in the past (e.g., Costello et al., 2013). Formalized data review has proved difficult, though, in part because of the heterogeneity of research data both across and within subdisciplines. Moreover, the ways of dealing with heterogeneity vary even within organizations. Mayernik (2016: 973) illustrated the complexity of developing standardized practices for data management and archiving through a case study of three programs, including LTER. The author found that "institutional support for data and metadata management are not uniform within a single organization or academic discipline." Vanderbilt et al. (2009) noted similar challenges in a study of data integration in the International LTER program, and the general issue of "science friction" is well-documented in social studies of science (e.g., Edwards et al., 2011).

To be sure, an important distinction exists between data curation efforts—which seek to organize, integrate, and preserve datasets for long-term sustainability and reuse for various purposes—and peer review efforts, which seek to ensure that data analysis underlying manuscript findings reflect scientific best-practices and that datasets are sufficiently complete to evaluate said findings.

Various examples of successful data curation efforts demonstrate that data review is one part of broader curation tasks and can perhaps be more easily standardized than data review for academic manuscripts. Standardization is perhaps more achievable for data curation efforts because funding agencies can issue strong mandates and sufficient funding for data review. One such example is the NSF Arctic Data Center's process for requiring "metadata, full data sets, and derived data products be deposited in a long-lived and publicly accessible archive" (NSF, 2016). Data review processes for journals, on the other hand, must be flexible enough to accommodate heterogeneity and complexity of datasets and the institutional structures in which data are generated while achieving some degree of standardization.

Data librarians and other research support staff comprise a community that is familiar with striking such a balance. As Lin and Strasser (2014) noted, "librarians, information technologists, preservation specialists, and others have a long history of providing infrastructure, education, and support for preserving and promoting researchers' outputs," and they argue that scholarly publishers should elevate that role for the professionals listed. Journals might begin to do so by engaging with communities such as the Research Data Alliance (RDA) to develop data review processes. RDA and other groups engage regularly with public and private funders (Berman et al., 2014) and therefore have experience translating policy into the research process; however, journals appear to be less engaged with RDA and similar groups. The professionals in data management roles, including in non-library organizations (e.g., DataONE) might alleviate journals' difficulties navigating issues such as resource constriction for managing research outputs; the nuances of ownership rights and intellectual property; and principles of reuse, credit, and citation.

Journal editors frequently reported that the human resource cost of reviewing datasets presented a barrier to implementing data review processes. One option often discussed in the research data management community involves automating the data review process, which would alleviate some of the concern over reviewers' unpaid time and effort. Organizations such as the U.S. Geological Survey (2016) have outlined options for automating audits of data management practices to ensure adherence to best-practices. Building best-practices, as defined by individual communities (Moore et al., 2010) and by the broader research community (Tenopir et al., 2011; Wilson et al., 2014) into the research process from the outset of a project might aid in easing the burden of PDA on reviewers, editors, and authors at the manuscript stage. Automation tools provide a starting point for ensuring that researchers include appropriate metadata, handle missing data with care, control versions, and format data in machine-readable formats (Starr et al., 2015).

### 5.1.2 Applying strategies from software review initiatives

Our findings also illustrated some carryover effects from editors' experiences with PDA policies into their perceptions of impending code review and publication policies. In principle, there is little reason to treat data archiving and code archiving serially; the two research products are often intertwined, but also have meaningful differences that require attention. In practice, though, journals are beginning to draft and implement code sharing policies by extending existing policies or issuing new ones. It is worthwhile, then, to consider how journals might simultaneously (1) apply mechanisms developed by journals and organizations who regularly facilitate peer review of *software* (distinct from code, to be discussed below) to data review and (2) adapt these processes to future code review and archiving initiatives.

The perception among editors appears to be that scientists will be unprepared to share code used to analyze data due to inexperience with code sharing best-practices (e.g., adequate documentation, appropriate use of versioning technologies), concerns that mirror the issues editors raised regarding data sharing. Perhaps as a result, editors described potential code archiving policies as almost identical to open data policies in function: mandated availability, but little attention to how the review process should incorporate code review. Rather than framing editors' conflation of data and code review and archiving as a barrier to developing sound policies, we propose that the development of code sharing policies may help to improve journal data policies and their application.

Consider, for example, the similarities in challenges that both datasets and code present when compared to manuscripts. Both artifacts can be heterogeneous in format, have traditionally been under-supported and/or underappreciated aspects of the research process, and are increasingly reliant on computing infrastructures. Furthermore, editors expressed concern about finding volunteer reviewers with expertise in data and code in addition to scientific knowledge on a manuscript's subject matter. Various organizations and initiatives have begun to tackle these issues as they relate to software and may provide a foundation for improving journal policies for both data and code sharing.

Open source software (OSS) journals and other OSS organizations hold collective lessons to offer journals in the way of mobilizing expertise to review data and code before archiving. These organizations employ processes similar to peer review to ensure that software—or standalone, reusable packages of code for various types of scientific computing—adhere to best-practices and community standards. Software is distinct from the code that might accompany a manuscript in that it typically can be used for more general purposes (e.g., data retrieval, database access, or text analysis)[1] than analysis on a single dataset for a single study (Barnes, 2010). Software review employed by OSS organizations and specialized journals such as *Methods in Ecology and Evolution* (in the case of software papers) includes code review in addition to a number of other checks, including dependencies on other packages and appropriate licensing.

---
[1] See rOpenSci Onboarding Policies: https://ropensci.github.io/dev_guide/policies.html

Just as previous authors have suggested that government agencies look to open source culture for lessons on how to develop their own open data initiatives (See Baack, 2015; Schrock and Shaffer, 2017), journals might also translate open source processes into peer review. Despite the distinction between code and software described above, journals might look to OSS organizations for guidance and partnership in reviewing code just as they might work with research data specialists on PDA policies. For example, the *Journal of Open Source Software* (see Smith et al., 2018), the *Journal of Open Research Software*, *SoftwareX*, and a host of discipline-specific journals (Chue Hong, n.d.) have generally converged on guidelines for reviewers when examining research software. Adopting or adapting these practices—whether formally or informally—might aid non-software journals in educating reviewers and developing mechanisms for code review and archiving.

The formal option is a partnership between journals and OSS organizations, a model that might be applied to both data and code review. These partnerships are already emerging at the intersection of ecology research and software. For example, rOpenSci's partnership with *Methods in Ecology & Evolution* enables authors to submit the software used in manuscript production to rOpenSci's review process (Methods.blog, 2017). The review ensures that software packages adhere to best-practices and offers authors the benefit of package promotion, support and maintenance, and community feedback,[2] incentives that may promote quality improvements in research products. OSS organizations and other journals might seek similar, mutually beneficial relationships. Journals would benefit from increased submissions by also offering an additional process to house datasets and software packages in high-visibility locations. OSS organizations, in return, expand their reach and aggregate useful metrics of impact, such as citation count. Furthermore, OSS organizations and journals alike tend to seek partnerships that extend an existing capability, making OSS' existing processes for scrutinizing data and code a match for journals' existing peer review processes.

## 6. CONCLUSION

In summary, journals in ecology and evolution have made strides in ensuring the availability and sustainability of datasets by instituting policies that go beyond "available upon request." However, challenges remain in developing mechanisms for ensuring that the archived data are complete and useful for the outcomes archiving policies wish to achieve. Trusting authors to archive complete and high-quality datasets is problematic in that authors may or may not comply for various reasons. For instance, some may fear "scooping" of their projects or feel that the time cost of cleaning datasets for sharing outweighs potential benefits.

---

[2] See rOpenSci "Why submit your package to rOpenSci?" for an example: https://ropensci.github.io/dev_guide/onboardingintro.html#whysubmit

Additionally, if journals continue to rely on hope in reviewers and trust in authors for quality data submissions, we might expect change to occur slowly. A few ways to accelerate the process include learning from data curation initiatives and software review mechanisms used by OSS organizations, which are beginning to solidify in the digital curation community, software journals and, to a lesser extent, in disciplinary journals in ecology and evolution.

Journals will constitute a valuable organization for studying the governance of data and software sharing and comparing between them, across disciplines, and in seemingly infinite other configurations because they are ubiquitous in the process of publishing outputs of scientific research. Studying the experiences of journal editors, reviewers, and authors during the rollout of code sharing policies can help open source software organizations and others reflect on management principles/policies and strategies for organizing.

## 7. ACKNOWLEDGMENTS

We would like to thank the interview participants who graciously volunteered their time to be interviewed for our study. Research reported in this publication was supported in whole by Leona M. and Harry B. Helmsley Charitable Trust. Research was conducted under UC Berkeley Office for the Protection of Human Subjects Protocol ID 2017-08-10194.

## 8. REFERENCES

Baack S (2015) Datafication and empowerment: How the open data movement re-articulates notions of democracy, participation, and journalism. *Big Data & Society* 2(2): p.2053951715594634.

Barnes N (2010) Publish your computer code: it is good enough. 467(7317): 753.

Beardsley TM (2010) The biologist's burden. *BioScience* 60(7): 483-484.

Berman F, Wilkinson R and Wood J (2014) Guest editorial: Building global infrastructure for data sharing and exchange through the Research Data Alliance. *D-Lib Magazine* 20(1/2). doi:10.1045/january2014-berman

Bernard HR (1988) *Research in Cultural Anthropology*. Thousand Oaks: Sage Publishers.

Bloom T, Ganley E and Winker M (2014) Data access for the open access literature: PLOS's data policy. *PLoS Biology* 12(2): e1001797. https://doi.org/10.1371/journal.pbio.1001797

Borgman CL (2007) *Scholarship in the Digital Age: Information, Infrastructure, and the Internet.* Cambridge, MA: MIT Press.

Bowker GC (2001) Biodiversity datadiversity. *Social Studies of Science* 30(5): 643–684.

Bowker GC (2006) *Memory Practices in the Sciences.* Cambridge, MA: MIT Press.

Chavan VS and Ingwersen P (2009) Towards a data publishing framework for primary biodiversity data: challenges and potentials for the biodiversity informatics community. *BMC Bioinformatics* 10(14): S2. https://doi.org/10.1186/1471-2105-10-S14-S2

Chue Hong N (n.d.) In which journals should I publish my software? Software Sustainability Institute. Available at: https://www.software.ac.uk/resources/guides/which-journals-should-i-publish-my-software (accessed 01 September 2018).

Coleman DC (2010) *Big Ecology: The Emergence of Ecosystem Science.* Berkeley, CA: Univ of California Press.

Corbin JM and Strauss A (1990) Grounded theory research: Procedures, canons, and evaluative criteria. *Qualitative Sociology* 13(1): 3-21.

Costello MJ (2009) Motivating online publication of data. *BioScience* 59(5): 418-27.

Costello MJ, Michener WK, Gahegan M, Zhang ZQ and Bourne PE (2013) Biodiversity data should be published, cited, and peer reviewed. *Trends in Ecology & Evolution* 28(8): 454-61.

Couture JL, Blake RE, McDonald G and Ward CL (2018) A funder-imposed data publication requirement seldom inspired data sharing. *PloS One* 13(7): p.e0199789. https://doi.org/10.1371/journal.pone.0199789

Diekema AR, Wesolek A and Walters CD (2014) The NSF/NIH Effect: Surveying the effect of data management requirements on faculty, sponsored programs, and institutional repositories. *The Journal of Academic Librarianship* 40(3-4): 322-331.

Dryad (2007) Workshop May 2007 ideas. Available at:
http://wiki.datadryad.org/Workshop_May_2007_Ideas (accessed 08 August
2018).

Duke CS and Porter JH (2013) The ethics of data sharing and reuse in biology.
*BioScience.* 63(6): 483-489.

Ecological Applications (n.d.) Data policy. Available at:
https://esajournals.onlinelibrary.wiley.com/hub/journal/19395582/resources/da
ta-policy-eap (accessed 04 August 2018).

Ecologocial Society of America (n.d.) Ecological Archives. Available at:
http://esapubs.org/archive/ (accessed 04 August 2018).

Edwards PN, Mayernik MS, Batcheller AL, et al. (2011) Science friction: Data,
metadata, and collaboration. *Social Studies of Science* 41(5): 667–690.

Emerson RM, Fretz RI and Shaw LL (1995) *Writing Ethnographic Fieldnotes.* Chicago:
University of Chicago Press.

Fairbairn DJ (2011) The advent of mandatory data archiving. *Evolution* 65(1): 1-2.

Genome Research (n.d.) Material and data release policy for papers published in
*Genome Research.* Available at:
https://genome.cshlp.org/site/misc/matdatarel.xhtml (accessed 28 August 2018).

Glaser BG (1965) The constant comparative method of qualitative analysis. *Social
Problems* 12(4): 436-445.

Goecks J, Nekrutenko A, and Taylor J (2010) Galaxy: a comprehensive approach for
supporting accessible, reproducible, and transparent computational research in the
life sciences. *Genome Biology* 11(8): R86. https://doi.org/10.1186/gb-2010-11-8-r86

Group on Earth Observations (GEO) (2015) *The Value of Open Data Sharing.* Paris:
Committee on Data for Science and Technology (CODATA). 42 p. Available at:
https://www.earthobservations.org/documents/dsp/20151130_the_value_of_op
en_data_sharing.pdf (accessed 02 August 2018).

Hilgartner S (2013) Constituting large-scale biology: Building a regime of
governance in the early years of the Human Genome Project. *BioSocieties* 8(4): 397-
416.

Hine C (2006) Databases as scientific instruments and their role in the ordering of scientific work. *Social Studies of Science* 36(2): 269–298

Jackson SJ, Edwards PN, Bowker GC and Knobel CP (2007) Understanding infrastructure: History, heuristics and cyberinfrastructure policy. *First Monday* 2007 12(6). http://firstmonday.org/issues/issue12_6/jackson/index.html

Jasanoff S (Ed.) (2004) *States of Knowledge: The Co-production of Science and Social Order.* New York: Routledge.

Kaye J and Hawkins N (2014) Data sharing policy design for consortia: challenges for sustainability. *Genome Medicine* 6(1): 4.

Knorr Cetina K (1998) *Epistemic Cultures: How the Sciences Make Knowledge.* Cambridge, MA: Harvard University Press.

Leonelli S (2014) What difference does quantity make? On the epistemology of Big Data in biology. *Big Data & Society* 1(1): p.2053951714534395.

Lin J and Strasser C (2014) Recommendations for the role of publishers in access to data. *PLoS Biology* 12(10): e1001975. https://doi.org/10.1371/journal.pbio.1001975

Long Term Ecological Research Network (LTER) (2018) LTER history. Available at: https://lternet.edu/network-organization/lter-a-history/2018 (accessed 05 August 2018).

Mayernik MS. Research data and metadata curation as institutional issues. *Journal of the Association for Information Science and Technology* 67(4): 973-93.

Mayernik MS (2017) Open data: Accountability and transparency. *Big Data & Society* 4(2): p.2053951717718853.

Methods.blog (2017) Software review collaboration with rOpenSci. *British Ecological Society.* Available at: https://methodsblog.wordpress.com/2017/11/29/software-review/ (accessed 23 August 2018).

Michener WK (2015) Ecological data sharing. *Ecological Informatics* 29: 33-44.

Michener WK and Jones MB (2012) Ecoinformatics: supporting ecology as a data-intensive science. *Trends in Ecology & Evolution* 27(2): 85-93.

Miles MB and Huberman AM (1994) *Qualitative Data Analysis: An Expanded Sourcebook.* Thousand Oaks: Sage Publishers.

Moore AJ, McPeek MA, Rausher MD, Rieseberg L and Whitlock MC (2010) The need for archiving data in evolutionary biology. *Journal of Evolutionary Biology* 23(4): 659-60.

National Science Foundation (NSF) (2016) Dear colleague letter: Data management and data reporting requirements for research awards supported by the Office of Polar Programs. NSF 16-055. Available from: https://www.nsf.gov/pubs/2016/nsf16055/nsf16055.jsp (accessed 05 August 2018).

Nelson B (2009) Data sharing: Empty archives. *Nature News* 461(7261): 160-163.

Nielsen M (2011) *Reinventing Discovery: The New Era of Networked Science.* Princeton: Princeton University Press.

Piwowar HA, Day RS and Fridsma DB (2007) Sharing detailed research data is associated with increased citation rate. *PloS One* 2(3): e308. https://doi.org/10.1371/journal.pone.0000308

PLoS ONE (n.d.) Data availability. Available at: https://journals.plos.org/plosone/s/data-availability (accessed 14 August 2018).

Roche DG, Kruuk LE, Lanfear R and Binning SA (2015) Public data archiving in ecology and evolution: how well are we doing? *PLoS Biology* 13(11): e1002295. https://doi.org/10.1371/journal.pbio.1002295

The Royal Society Science Policy Centre. *Science as an Open Enterprise.* London: The Royal Society. Available at: https://royalsociety.org/~/media/policy/projects/sape/2012-06-20-saoe.pdf (accessed 10 August 2018).

rOpenSci (n.d.) Onboarding policies. Available at: https://ropensci.github.io/dev_guide/policies.html (accessed 10 August 2018).

rOpenSci (n.d.) Why submit your package to rOpenSci? Available at: https://ropensci.github.io/dev_guide/onboardingintro.html#whysubmit (accessed 10 August 2018).

Scientific Data (n.d.) For authors. Available at: https://www.nature.com/sdata/publish/for-authors#aims-scope (accessed 11 August 2018).

Science Europe (2018) *Science Europe's Framework for Discipline-specific Research Data Management*. Available at: https://www.scienceeurope.org/wp-content/uploads/2018/01/SE_Guidance_Document_RDMPs.pdf (accessed 08 August 2018).

Schrock A and Shaffer G (2017) Data ideologies of an interested public: A study of grassroots open government data intermediaries. *Big Data & Society* 4(1): p.2053951717690750.

South DB and Duke CS (2010) Will a data registry increase professional integrity? *Journal of Forestry*. 108(7): 370-371.

Spradley JP (1979) *The Ethnographic Interview*. New York: Waveland Press.

Star SL and Ruhleder K (1996) Steps toward an ecology of infrastructure: Design and access for large information spaces. *Information Systems Research* 7(1): 111-34.

Starr J, Castro E, Crosas M, Dumontier M, Downs RR, Duerr R, Haak LL, Haendel M, Herman I, Hodson S and Hourclé J et al. (2015) Achieving human and machine accessibility of cited data in scholarly publications. *PeerJ Computer Science* 1:e1. http://doi.org/10.7717/peerj-cs.1

Stodden V, Seiler J and Ma Z (2018) An empirical analysis of journal policy effectiveness for computational reproducibility. *Proceedings of the National Academy of Sciences* 115(11): 2584-2589.

Tenopir C, Allard S, Douglass K, Aydinoglu AU, Wu L, Read E, Manoff M and Frame M (2011) Data sharing by scientists: practices and perceptions. *PloS One* 6(6): e21101. https://doi.org/10.1371/journal.pone.0021101

Tilson D, Lyytinen K and Sørensen C. (2010) Research commentary—Digital infrastructures: The missing IS research agenda. *Information Systems Research* 21(4): 748-759.

United States Geological Survey (USGS) (2016) Data management plan solution comparison charts. Available at: https://my.usgs.gov/confluence/download/attachments/552932046/DMPComparisonsChart.pdf?version=1&modificationDate=1472225589700&api=v (accessed 01 September 2018).

Vanderbilt K, Cushing J, Gao J, Kaplan N, Kruger J, Leroy C, Mallett J, Ramsey K and Zeman L (2009) Data integration challenges: an example from the International Long-Term Ecological Research Network (ILTER). *Ecological Circuits*. 2: 12-13.

Vines TH, Albert AY, Andrew RL, Débarre F, Bock DG, Franklin MT, Gilbert KJ, Moore JS, Renaut S and Rennison DJ (2014) The availability of research data declines rapidly with article age. *Current Biology* 24(1): 94-97.

Vines TH, Andrew RL, Bock DG, Franklin MT, Gilbert KJ, Kane NC, Moore JS, Moyers BT, Renaut S, Rennison DJ and Veen T (2013) Mandated data archiving greatly improves access to research data. *The FASEB Journal* 27(4): 1304-1308.

Ware M and Mabe M (2015) *The STM Report: An Overview of Scientific and Scholarly Journal Publishing*. Den Haag: International Association of Scientific, Technical and Medical Publishers. Available at: https://digitalcommons.unl.edu/scholcom/9/ (accessed 08 August 2018).

Weiss RS (1995) *Learning from Strangers: The Art and Method of Qualitative Interview Studies*. New York: Simon and Schuster.

Whitlock MC (2011) Data archiving in ecology and evolution: best practices. *Trends in Ecology & Evolution* 26(2): 61-65.

Whitlock MC, McPeek MA, Rausher MD, Rieseberg L and Moore AJ. Data archiving. *The American Naturalist* 175(2): 145-146.

Wilkinson MD, Dumontier M, Aalbersberg IJ, Appleton G, Axton M, Baak A, Blomberg N, Boiten JW, Da Silva Santos LB, Bourne PE and Bouwman J. The FAIR Guiding Principles for scientific data management and stewardship. *Scientific Data* 3: 1-9.

Wilson G, Aruliah DA, Brown CT, Chue Hong NP, Davis M, Guy RT, Haddock SH, Huff KD, Mitchell IM, Plumbley MD and Waugh B (2014) Best practices for scientific computing. *PLoS Biology* 12(1): e1001745. https://doi.org/10.1371/journal.pbio.1001745

Yin RK (1994) *Case Study Research: Design and Methods*. Beverly Hills: Sage Publications.

## APPENDIX A - Journals with Data Archiving Policies

From https://en.wikipedia.org/wiki/Research_data_archiving:

- *The American Naturalist*
- *Biological Journal of the Linnean Society*
- *Biology Letters*
- *BMC Ecology*
- *BMC Evolutionary Biology*
- *BMJ Open*
- *Ecological Applications and Ecological Monographs*
- *Evolution*
- *Evolutionary Applications*
- *Functional Ecology*
- *Genetics*
- *Heredity*
- *Journal of Applied Ecology*
- *Journal of Ecology*
- *Journal of Evolutionary Biology*
- *Journal of Fish and Wildlife Management*
- *Journal of Heredity*
- *Journal of Paleontology*
- *Methods in Ecology and Evolution*
- *Molecular Ecology*
- *Molecular Ecology Resources*
- *Nature*
- *Nucleic Acids Research*
- *Paleobiology*
- *PLOS*
- *Science*
- *Systematic Biology*